\documentclass[%
reprint,
superscriptaddress,
showpacs,
showpacs,preprintnumbers,
amsmath,amssymb,
aps,
]{revtex4-1}

\usepackage[dvipdfmx]{graphicx}
\usepackage{dcolumn}
\usepackage{bm}
\usepackage{multirow}
\usepackage[mathlines]{lineno}
\usepackage{comment}

\usepackage{color}

\begin{document}

\preprint{APS/123-QED}

\title{Successive destruction of charge density wave states by pressure in LaAgSb$_2$}

\author{Kazuto~Akiba}
\email{akb@okayama-u.ac.jp}
\affiliation{
Graduate School of Natural Science and Technology, 
Okayama University, Okayama 700-8530, Japan
}

\author{Hiroaki~Nishimori}
\affiliation{
Graduate School of Natural Science and Technology, 
Okayama University, Okayama 700-8530, Japan
}

\author{Nobuaki~Umeshita}
\affiliation{
Graduate School of Natural Science and Technology, 
Okayama University, Okayama 700-8530, Japan
}

\author{Tatsuo~C.~Kobayashi}
\affiliation{
Graduate School of Natural Science and Technology, 
Okayama University, Okayama 700-8530, Japan
}

\date{\today}

\begin{abstract}
We comprehensively studied the magnetotransport properties of LaAgSb$_2$ under high pressure up to 4 GPa,
which showed unique successive charge density wave (CDW) transitions at $T_{CDW1}\sim 210$ K
and $T_{CDW2}\sim 190$ K at ambient pressure.
With the application of pressure, both $T_{CDW1}$ and $T_{CDW2}$ were suppressed and
disappeared at the critical pressures of $P_{CDW1}=3.0$--3.4 GPa and $P_{CDW2}=1.5$--1.9 GPa, respectively.
At $P_{CDW1}$, the Hall conductivity showed a step-like increase, which is consistently understood by the emergence
of two-dimensional hollow Fermi surface at $P_{CDW1}$.
We also observed a significant negative magnetoresistance effect when
the magnetic field and current were applied parallel to the $c$ axis.
The negative contribution was observed in whole pressure region from 0 to 4 GPa.
Shubnikov--de Haas (SdH) oscillation measurements under pressure directly showed the changes in the Fermi surface across the CDW phase boundaries.
In $P<P_{CDW2}$, three major oscillation components, $\alpha$, $\beta$, and $\gamma$, were identified, whose frequencies were increased by application of pressure.
The increment rate of these frequencies was considerably larger than that expected from the shrinkage of lattice constant, indicating the unignorable
band modification under pressure.
In the normal metallic phase above $P>P_{CDW1}$,
we observed a single frequency of $\sim 48$ T with a cyclotron effective mass of 0.066 $m_0$,
whose cross section in the reciprocal space corresponded to only 0.22\% of the first Brillouin zone.
Besides, we observed another oscillation component with frequency of $\sim 9.2$ T,
which is significantly enhanced in the limited pressure range of $P_{CDW2}<P<P_{CDW1}$.
The amplitude of this oscillation was anomalously suppressed in the high-field and
low-temperature region, which cannot be explained by the conventional Lifshitz--Kosevich formula.
\end{abstract}

\maketitle

\section{Introduction}
The charge density wave (CDW) has long been known as a macroscopic quantum state, which shows a static order of charge carrier accompanied by lattice modulation.
Several three-dimensional materials with anisotropic crystal structure and Fermi surface show spontaneous formation of a CDW state with a lattice modulation wavenumber of $\bm{Q}$ by lowering the temperature \cite{Gruner_CDW}.
Following the Peierls transition in one-dimensional system,
the mechanism of CDW is frequently understood as a nesting of Fermi surfaces connected by
a nesting vector $\bm{Q}$, which results in an opening of the energy gap at the the nested region.
The vanishing of the corresponding Fermi surface considerably alters the density of state at the Fermi level and
causes drastic changes in various physical properties.
In real three-dimensional material with multiple Fermi surfaces, however,
it is a long-standing issue under intense discussion whether the formation of CDW can be understood fully with only the Fermi surface properties 
\cite{Johannes_2008, Eiter_2013}.

\begin{figure}[]
\centering
\includegraphics[]{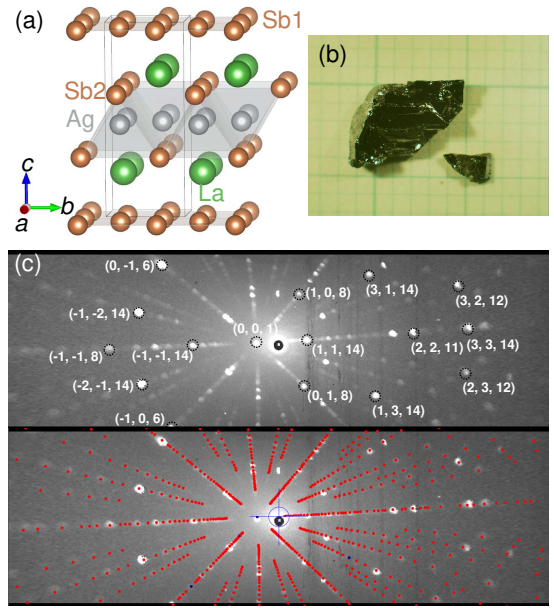}
\caption{
(a) Crystal structure of LaAgSb$_2$. Shaded surfaces represent the square net formed by Sb1 and the tetrahedron formed by Sb2.
(b) A picture of LaAgSb$_2$ single crystal utilized in the present study.
The smallest square in the background is 1mm $\times$ 1 mm.
(c) Back-reflection Laue pattern of LaAgSb$_2$ single crystal with X-ray beam parallel to the $c$ axis.
The lower panel shows the simulated pattern using crystallographic data obtained by single-crystal structural analysis.
\label{fig_1}}
\end{figure}

The layered intermetallic compound LaAgSb$_2$, the target of the present study,
forms a tetragonal crystal structure ($P4/nmm$, space group No. 129) \cite{Brylak_1995}, as
shown in Fig. \ref{fig_1}(a).
Ag and Sb forms a layered structure similar to the iron-based superconductors \cite{Johrendt_2011},
in which Sb1 forms square nets
and Ag is located in a slightly distorted tetrahedron formed by Sb2.
La is located between these layers in an alternating manner along the $c$ axis.

LaAgSb$_2$ has been known to exhibit successive CDW transitions
at $T_{CDW1}\sim 210$ K and $T_{CDW2}\sim 190$ K.
An X-ray diffraction study \cite{Song_2003} revealed that the CDW1 has a lattice modulation along the $a$ axis
characterized by modulation vector ($0.026\times 2\pi/a$, 0, 0),
while the CDW2 has a modulation along the $c$ axis with modulation vector (0, 0, $0.16\times 2\pi/c$),
where $a$ and $c$ represent the lattice constants.
Both CDW1 and CDW2 have been explained as a result of Fermi-surface nesting \cite{Song_2003}.
The phase transition to the CDW1 can be observed as a clear hump-like anomaly in the temperature dependence of the resistivity \cite{Myers_1999a}.
Compared to CDW1, the phase transition to CDW2 causes a smaller anomaly in the electrical transport, which is barely observed in the out-of-plane resistivity \cite{Song_2003, Watanabe_2006}.
This difference has been explained as a larger segment of the Fermi surface being gapped by the CDW1 transition,
which causes a significant change in the carrier density.

The Fermi surfaces of LaAgSb$_2$ have been studied by quantum oscillation measurements \cite{Myers_1999b, Inada_2002, Budko_2008} combined with first principles calculations and angle-resolved photoemission spectroscopy (ARPES) measurements \cite{Arakane_2006, Shi_2016}.
Although several controversial conclusions can be seen in previous literature
(e.g., regarding the shape of Fermi surfaces and
correspondence of Fermi surface cross sections with quantum-oscillation frequencies),
LaAgSb$_2$ is considered to have four Fermi surfaces:
two-hole Fermi surfaces around the $\Gamma$ point,
a two-dimensional hollow electron surface, and an ellipsoidal electron surface around the $X$ point \cite{Myers_1999b}.
Intriguingly, a recent ARPES measurement reported that the hollow Fermi surface, which is considered responsible for the nesting of CDW1,
hosts a Dirac-cone-like dispersion in the vicinity of the Fermi level \cite{Shi_2016}.
Although a part of this linear dispersion is considered to be gapped by the formation of CDW1 at ambient pressure,
the large linear magnetoresistance that is frequently observed in Dirac fermion systems 
and the existence of high mobility carrier have been reported in a magnetotransport study \cite{Wang_2012}.
If we can eliminate the CDW1 by tuning an external parameter,
we can expect the full contribution of the Dirac-like dispersion that is hidden by the CDW energy gap at ambient pressure.
Thus, LaAgSb$_2$ can be a typical example
to investigate the exotic transport phenomena originating from the Dirac fermion and the interplay between CDW formation and the presence of the Dirac fermion.

Replacement of the Ag site with its homologues, Au and Cu, has been known to cause a drastic change in the CDW properties.
In LaCuSb$_2$, the temperature dependence of the resistivity is monotonic in contrast with LaAgSb$_2$ \cite{Chamorro_2019},
and no phase transition has been reported to date.
However, LaAuSb$_2$ shows CDW transition with clear hump-like anomaly at $T_{CDW1}\sim$100 K in resistivity \cite{Seo_2012, Kuo_2019}, as well as LaAgSb$_2$,
though the $T_{CDW1}$ is considerably lower than that of LaAgSb$_2$.
The dimensionality of the Fermi surfaces is considered a possible factor to determine the occurrence of CDW \cite{Hase_2014},
though the specific origin of this difference is not clear at the present stage.
Interestingly, in a recent study of LaAuSb$_2$, it has been reported that there exists a secondary CDW transition
at $T_{CDW2}\sim90$ K at ambient pressure \cite{Xiang_2020},
suggesting that LaAuSb$_2$ and LaAgSb$_2$ have quite similar CDW phase diagrams.
Thus, it is important to compare and contrast the CDW properties between LaAgSb$_2$ and LaAuSb$_2$
to understand the origin of the CDW formation in this material class.
In case of LaAuSb$_2$, both $T_{CDW1}$ and $T_{CDW2}$ are suppressed by application of pressure, and disappear at 1.7 GPa and 0.75 GPa , respectively \cite{Xiang_2020}.

The application of pressure and chemical substitution of La/Ag sites are known as effective methods to control the $T_{CDW1}$ and $T_{CDW2}$ in LaAgSb$_2$ \cite{Budko_2006, Torikachvili_2007, Masubuchi_2014}.
However, chemical substitution usually introduces lattice defects and unintentional carrier doping,
which hinders systematic investigation of the CDW states.
In contrast, high pressure is a useful external parameter that is free from the problems mentioned above.
However, the effect of pressure on the electronic state has only been studied below 2.1 GPa \cite{Budko_2006, Torikachvili_2007},
which is insufficient to eliminate the CDW1.
At present,
the critical pressures of the CDWs in LaAgSb$_2$ has remained unclear.

To clarify the complete pressure--temperature phase diagram of CDWs and to investigate the change in Fermi surfaces by application of pressure,
we comprehensively studied magnetotransport properties of LaAgSb$_2$ under high pressure up to 4 GPa.

\section{Experimental methods}
Single crystals of LaAgSb$_2$ were obtained by the Sb self-flux method \cite{Myers_1999a}.
La (99.9\%), Ag (99.99\%), and Sb (99.9999\%) with a molar ratio of 1 : 2 : 20 were placed in an alumina crucible,
and sealed in a quartz ampoule with argon gas.
After the mixture was heated to $1150\ {}^\circ\mathrm{C}$, it dwelled at this temperature for 12 hours.
Then, it was cooled to $670\ {}^\circ\mathrm{C}$ for 110--120 h.
The flux was removed using a centrifuge separator.
A picture of the as-grown single crystal is shown in Fig. \ref{fig_1}(b).
Most of crystals obtained by above method were millimeter-size lumps or plates.
The crystals were appropriately shaped into rectangle for electrical resistivity measurements.

X-ray single-crystal structural analysis at ambient pressure and $T=240$ K was performed using VariMax with Saturn (RIGAKU) with monochromated Mo K$_\alpha$ radiation ($\lambda=0.71075$ \AA).
The obtained data collection are listed in Tab. \ref{tab:xtal_collection} and \ref{tab:xtal_structure}.
Laue diffraction measurement was performed using IPX-YGR (IPX Co., Ltd.) with back-reflection configuration.
We obtained quite clear Laue spots, as shown in Fig. \ref{fig_1}(c), which is explained by the identified crystallographic data and indicates high quality of our single crystal.

\begin{table}
\caption{\label{tab:xtal_collection}
Crystallographic data and refinement statistics for X-ray single-crystal structural analysis at ambient pressure and $T=240$ K.
}
\begin{ruledtabular}
\begin{tabular}{ll}
formula&LaAgSb$_2$\\
crystal system&tetragonal\\
space group&$P 4/n m m$ (\#129)\\
$a$ (\AA)&4.3941(18)\\
$c$ (\AA)&10.868(6)\\
volume (\AA$^3$)&209.84(17)\\
Z value&2\\
$R1$ [$I >2\sigma(I)$]&0.0302\\
\end{tabular}
\end{ruledtabular}
\end{table}

\begin{table}
\caption{\label{tab:xtal_structure}
Atomic coordinates ($x$, $y$, $z$) and equivalent isotropic atomic displacement parameters ($U_{eq}$) at ambient pressure and $T=240$ K.
}
\begin{ruledtabular}
\begin{tabular}{llllll}
atom&site&$x$&$y$&$z$&$U_{eq}$ (\AA$^2$)\\
\hline
La&2c&1/4&1/4&0.23969(10)&0.0095(3)\\
Sb1&2a&3/4&1/4&0&0.0110(4)\\
Sb2&2c&3/4&3/4&0.33036(11)&0.0102(4)\\
Ag&2b&3/4&1/4&1/2&0.0133(4)\\
\end{tabular}
\end{ruledtabular}
\end{table}

The electrical transport under high pressure were measured
by indenter-type pressure cell ($P < 4$ GPa) \cite{Kobayashi_2007}.
Daphne oil 7474 \cite{Murata_2008} was used as a pressure medium.
The pressure in the sample space was determined by the superconducting transition temperature of Pb set near the sample.

Temperature dependence of the resistivity at zero field was measured by using a gas-flow-type optical cryostat (Oxford Instruments, $T > 2$ K)
and by a standard four-terminal method with a 2400 sourcemeter and 2182A nanovoltmeter (Keithley Instruments).
The effect of thermal electromotive force by temperature gradient was removed by inversion of the current ($I$) direction.

The transverse magnetoresistivity $\rho_{xx}$, longitudinal magnetoresistivity $\rho_{zz}$, and Hall resistivity $\rho_{yx}$ in a static magnetic field were measured using PPMS (Quantum Design, $B < 9$ T and $T > 2$ K), or superconducting magnet with variable-temperature insert (Oxford Instruments, $B < 8$ T and $T > 1.6$ K).
For the latter system, LR-700 AC resistance bridge (Linear Research) were utilized with the measurement frequency of 16 Hz.
They were measured by the standard four-terminal method, and electrical contacts were formed by silver paste (Dupont 4922N).

\section{Results and discussion}
\begin{figure}[]
\centering
\includegraphics[]{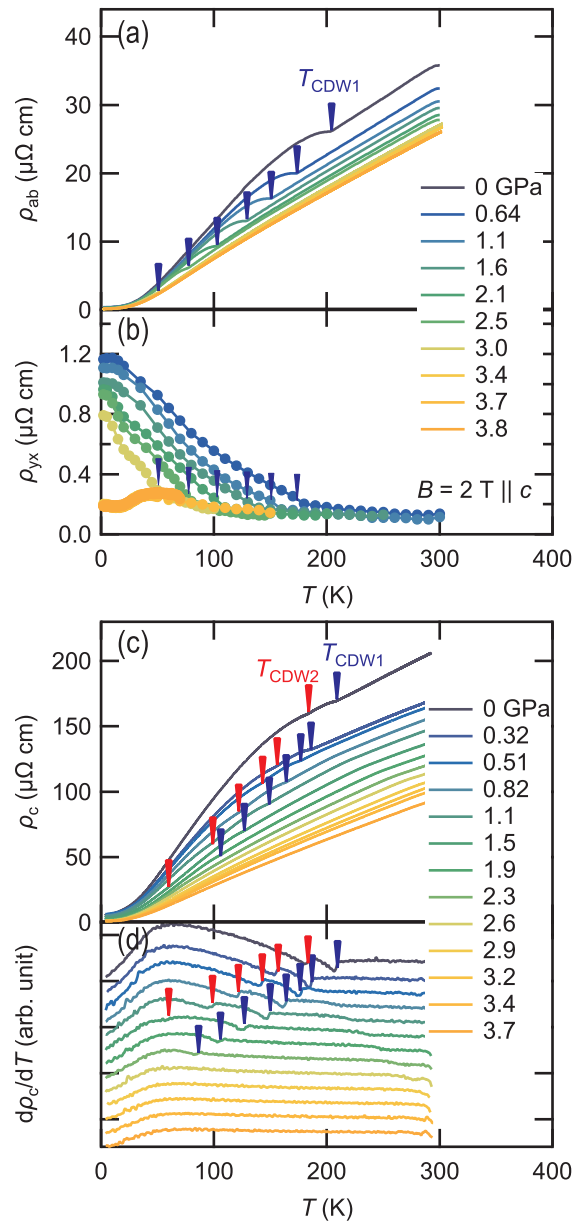}
\caption{
(a) Temperature dependence of the in-plane resistivity ($\rho_{ab}$) at several pressures.
Blue arrows indicate the transition temperature to CDW1 ($T_{CDW1}$).
(b) Temperature dependence of Hall resistivity $\rho_{yx}$ at $B=2$ T at several pressures.
(c) Temperature dependence of out-of-plane resistivity ($\rho_{c}$) at several pressures.
Red arrows indicate the transition temperature to CDW2 ($T_{CDW2}$).
(d) Temperature derivative of out-of-plane resistivity $d\rho_c/dT$ at several pressures.
The traces are vertically shifted for clarity.
\label{fig_2}}
\end{figure}
First, we focus on the temperature dependence of the in-plane resistivity ($\rho_{ab}$) at several pressures up to 3.8 GPa.
At ambient pressure, we observed an abrupt increase in $\rho_{ab}$ at 204 K, as indicated by a blue arrow in Fig. \ref{fig_2}(a), which is ascribed to $T_{CDW1}$.
This value agrees with the previous reports \cite{Masubuchi_2014, Budko_2006, Torikachvili_2007, Song_2003, Myers_1999a}.
With the increment in pressure, $T_{CDW1}$ monotonically decreased, and could not be defined above 3.4 GPa.
Complementary, the temperature dependence of Hall resistivity $\rho_{yx}$ at $B=2$ T is shown in Fig. \ref{fig_2}(b).
At all pressures below 3.0 GPa, a clear increase in $\rho_{yx}$ was observed at $T_{CDW1}$,
which may correspond to the disappearance of partial Fermi surfaces due to the transition to CDW1.
Above 3.4 GPa, however, $\rho_{yx}$ did not show an increase and hardly depended on pressure.
These results indicate that the critical pressure of CDW1 ($P_{CDW1}$) lies between 3.0 and 3.4 GPa.
$\rho_{yx}$ divided by the applied $B$ is the Hall coefficient $R_H$, which is represented by $R_H=1/[e(n_h-n_e)]$ in the high-field limit of the simple
electron--hole conduction model
with closed Fermi pockets.
Here, $n_h$, $n_e$, and $e$ represent the hole carrier density, electron carrier density, and elemental charge, respectively.
However, because the Fermi surface of LaAgSb$_2$ has complicated geometry,
$R_H$ cannot be connected simply with the carrier density, as discussed later.
Thus, we do not enter the quantitative analysis of $R_H$ in the present study.

As suggested by previous studies, an anomaly at $T_{CDW2}$ is quite subtle in $\rho_{ab}$, and we could not trace the pressure dependence of $T_{CDW2}$ in the in-plane transport.
Thus, we focused on the out-of-plane resistivity ($\rho_{c}$) to identify the $T_{CDW2}$.
Figure \ref{fig_2}(c) shows the temperature dependence of $\rho_{c}$ at several pressures.
$T_{CDW1}=209$ K is also clearly observed in $\rho_{c}$, whose pressure dependence is consistent with that in Figs. \ref{fig_2}(a) and (b).
Furthermore, we can identify a kink at 184 K at ambient pressure,
as indicated by a red arrow in Fig. \ref{fig_2}(c).
This is consistent with the transition temperature to CDW2 reported in the previous study \cite{Song_2003, Watanabe_2006}, and thus, we hereafter define this anomaly as $T_{CDW2}$.
As clearly observed in the $d\rho_c/dT$ in Fig. \ref{fig_2}(d), $T_{CDW2}$ was suppressed by pressure
and became invisible above 1.9 GPa.
From this result, we determined that the critical pressure of the CDW2 ($P_{CDW2}$) lies between 1.5 and 1.9 GPa.
\begin{figure}[]
\centering
\includegraphics[]{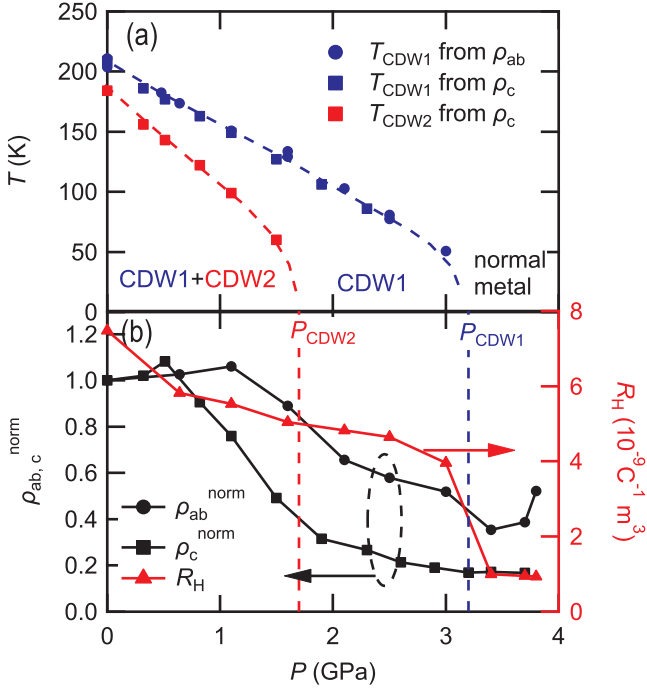}
\caption{
(a) Pressure--temperature phase diagram of LaAgSb$_2$.
The blue and red colors indicate $T_{CDW1}$ and $T_{CDW2}$, respectively,
and circular and rectangular markers indicate the transition temperature determined from $\rho_{ab}$ and $\rho_{c}$, respectively.
(b) Pressure dependences of the in-plane resistivity at 2 K ($\rho_{ab}^{norm}$),
out-of-plane resistivity at 3.7 K ($\rho_{c}^{norm}$),
and Hall coefficient ($R_H$).
$\rho_{ab}^{norm}$ and $\rho_{c}^{norm}$ were normalized by values at ambient pressure,
and $R_H$ was defined at $T=2$ K and $B =2$ T.
Vertical broken lines indicate the critical pressures of CDWs determined in the present study.
\label{fig_3}}
\end{figure}

We summarize the pressure dependence of $T_{CDW1}$ and $T_{CDW2}$ in Fig. \ref{fig_3}(a).
$T_{CDW1}$ determined from $\rho_{ab}$ and $\rho_{c}$ both lie on an identical line,
and the slope is determined to be $dT_{CDW1}/dP\sim -51$ K/GPa.
This value is slightly steeper than that reported in previous results ($-43$ K/GPa) \cite{Budko_2006, Torikachvili_2007}, but shows reasonable agreement.
Compared to $T_{CDW1}$, $T_{CDW2}$ is more rapidly suppressed by pressure with slope of $dT_{CDW2}/dP\sim -80$ K/GPa.
Both $T_{CDW1}$ and $T_{CDW2}$ depend linearly on the pressure sufficiently above $P_{CDW1}$ and $P_{CDW2}$,
whereas they seem to more rapidly approach zero in the vicinity of their critical pressure than in a simple linear extrapolation.
We note that $dT_{CDW1}/dP$ and $dT_{CDW2}/dP$ in LaAuSb$_2$ is estimated to be $-60$ and $-120$ K/GPa, respectively \cite{Xiang_2020},
suggesting that CDW orders in LaAgSb$_2$ is more robust against the application of pressure than those of LaAuSb$_2$.
The black traces in Fig. \ref{fig_3}(b) show the pressure dependences of the in-plane resistivity at 2 K ($\rho_{ab}^{norm}$) and out-of-plane resistivity at 3.7 K ($\rho_{c}^{norm}$) normalized by values at ambient pressure.
Up to $P_{CDW1}$,
$\rho_{ab}^{norm}$ shows decreasing trend.
In the normal metallic phase above $P_{CDW1}$, on the other hand, $\rho_{ab}^{norm}$ changes its slope and shows a rather weak pressure dependence.
$\rho_{c}^0$ also changes its slope at $P_{CDW2}$ and $P_{CDW1}$, and is almost constant in the normal phase.
A similar change in the slope at $P_{CDW1}$ is also reported in the pressure dependence of the in-plane residual resistivity of LaAuSb$_2$ \cite{Xiang_2020}. 
The red trace in Fig. \ref{fig_3}(b) shows $R_H$ defined by $\rho_{yx}$ at $B =2$ T,
which shows a significant decrease at $P_{CDW1}$.

\begin{figure}[]
\centering
\includegraphics[]{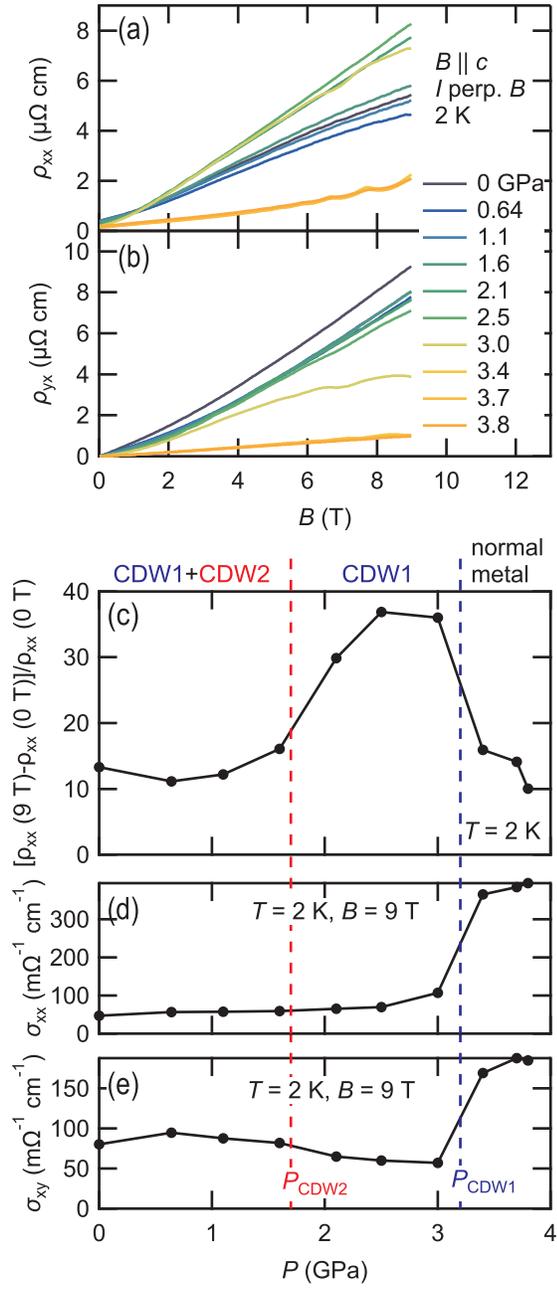}
\caption{
(a) In-plane magnetoresistivity ($\rho_{xx}$) and (b) Hall resistivity ($\rho_{yx}$) at 2 K at several pressures.
(c) Pressure dependence of magnetoresistivity at 2 K normalized by zero-field value [$\rho_{xx}$ (9 T)$-\rho_{xx}$ (0 T)]/$\rho_{xx}$ (0 T).
Pressure dependence of (d) $\sigma_{xx}$ and (e) $\sigma_{xy}$ at 2 K and 9 T.
\label{fig_4}}
\end{figure}
To obtain more insight on the change in electronic structure across $P_{CDW2}$ and $P_{CDW1}$,
we investigated the in-plane magnetoresistivity ($\rho_{xx}$) and Hall resistivity ($\rho_{yx}$) under high pressure with magnetic fields along the $c$ axis and electric current within the $a$--$b$ plane,
whose results are shown in Figs. \ref{fig_4}(a) and (b).
$\rho_{xx}$ at 2 K does not show the tendency of saturation and has quasilinear magnetic-field dependence rather than conventional quadratic one at all measured pressures.
$\rho_{yx}$ at ambient pressure shows almost identical property with the previous report \cite{Wang_2012}
and is significantly suppressed above 3.4 GPa.
In addition, we can see non-monotonic modulation superposed on $\rho_{xx}$ and $\rho_{yx}$, which is notably enhanced above 3.4 GPa.
This is ascribed to quantum oscillation, whose details will be discussed later.
As shown in Fig. \ref{fig_4}(c), the magnetoresistance effect shows characteristic behavior in each CDW phases.
Below $P_{CDW2}$, in which CDW1 and CDW2 coexist, the magnetoresistance effect at 9 T is approximately 10--20 and shows weak pressure dependence.
In the intermediate pressure range between $P_{CDW2}$ and $P_{CDW1}$, in which only CDW1 survives,
it is drastically enhanced up to 40. 
In the normal phase above $P_{CDW1}$, it rapidly decreases and again shows weak pressure dependence.

Using the data shown in Fig. \ref{fig_4}(a) and \ref{fig_4}(b), we deduced the pressure dependence of the in-plane electrical conductivity
$\sigma_{xx}=\rho_{xx}/(\rho_{xx}^2+\rho_{yx}^2)$ and
Hall conductivity $\sigma_{xy}=\rho_{yx}/(\rho_{xx}^2+\rho_{yx}^2)$ at $T=2$ K and $B = 9$ T, as shown in Fig. \ref{fig_4}(d) and (e).
Both $\sigma_{xx}$ and $\sigma_{xy}$ show step-like increase at $P_{CDW1}$, while there is no apparent change at $P_{CDW2}$.
$\sigma_{xx}$ in the normal metallic phase is approximately 8 times larger than that at ambient pressure,
indicating the realization of highly conductive phase above $P_{CDW1}$.
\begin{figure}[]
\centering
\includegraphics[]{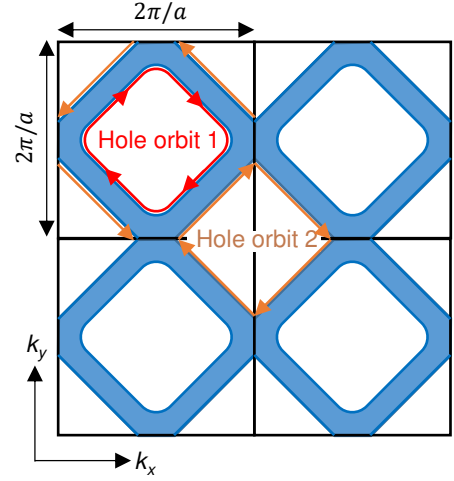}
\caption{
Schematic picture of two-dimensional hollow Fermi surface of LaAgSb$_2$ in the $k_x$--$k_y$ plane.
The black square represents the first Brillouin zone of the tetragonal cell.
The electron carrier occupies in the blue-shaded area.
Possible closed orbits are indicated by red and orange loops.
\label{orbit}}
\end{figure}
$\sigma_{xy}$ is represented by $\sigma_{xy}=(n_h-n_e)e/B$
in case of simple closed                                                                                                                                                                                                                                          
Fermi surface and $\omega_c \tau \gg 1$, where $\omega_c$ and $\tau$ represent the cyclotron frequency and relaxation time, respectively.
However, it should be noted that this picture cannot be applied directly in a complicated Fermi surface, as in LaAgSb$_2$.
More elementary, $\sigma_{xy}$ is represented by the following expression
with the cross sections of hole-like orbit $S_h(k_z)$ and electron-like orbit $S_e(k_z)$ cut by a constant-$k_z$ plane \cite{Abrikosov_FTM}:
\begin{equation}
\sigma_{xy}=\dfrac{e}{4\pi^3B}\int dk_z \left[S_h(k_z)-S_e(k_z)\right].
\end{equation}
Here, the hole-like and electron-like orbits represent the trajectories that enclose the higher- and lower-energy regions in the momentum space, respectively.
Let us consider a Fermi surface shown in Fig. \ref{orbit}, which is a simplified hollow Fermi surface of LaAgSb$_2$.
As mentioned above, the CDW1 is regarded to be result from a nesting within this Fermi surface.
In the first Brillouin zone, this Fermi surface hosts two types of closed orbit indicated by red and orange loops.
Because both orbits enclose a hollow region (high-energy region), these orbits should bring positive contribution to $\sigma_{xy}$.
This Fermi surface is more or less gapped out in the presence of CDW1; thus, the positive contribution to $\sigma_{xy}$ should be suppressed below $P_{CDW1}$.
Above $P_{CDW1}$, on the other hand, this Fermi surface fully contributes to the electrical transport, which results in enhancement of $\sigma_{xy}$.
This scenario qualitatively agree with the step-like increase in electrical conductivities at $P_{CDW1}$.
The absence of an anomaly at $P_{CDW2}$,
may suggest that the formation of CDW2 brings modest effect in the in-plane orbital motion.

\begin{figure}[]
\centering
\includegraphics[]{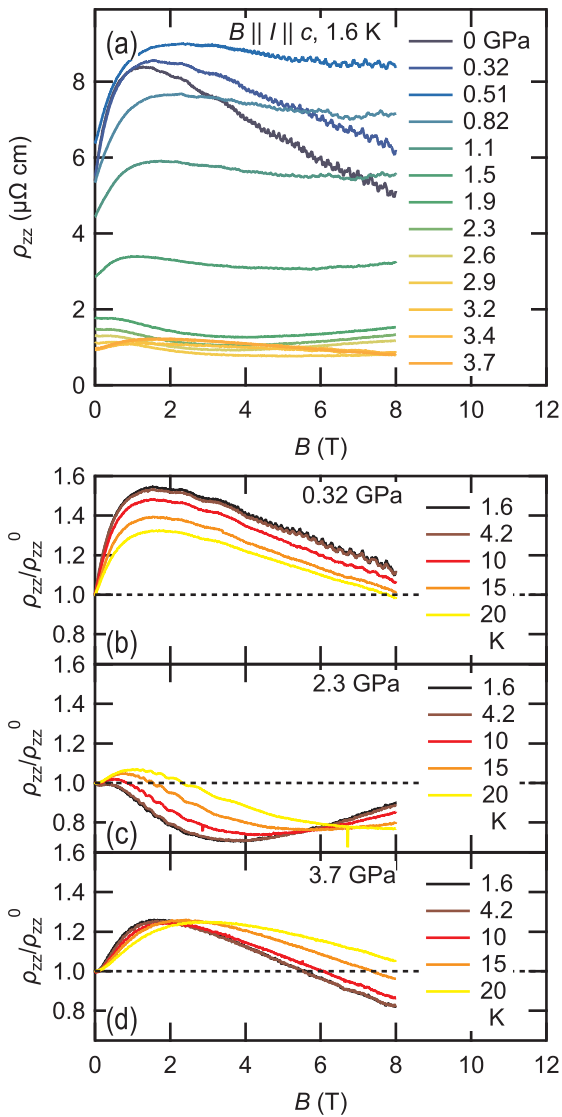}
\caption{
(a) Out-of-plane magnetoresistivity ($\rho_{zz}$) at 1.6 K at several pressures.
Temperature variation of out-of-plane magnetoresistivity normalized by zero-field value ($\rho_{zz}/\rho_{zz}^0$) at (b) 0.32, (c) 2.3, and (d) 3.7 GPa.
\label{fig_6}}
\end{figure}
We also investigated the out-of-plane magnetoresistivity ($\rho_{zz}$) at 1.6 K and up to 3.7 GPa, which is shown in Fig. \ref{fig_6}(a).
At ambient pressure, $\rho_{zz}$ rapidly increases as $B$ increases and has local maximum at approximately 1 T.
In higher $B$, we observed notable negative magnetoresistance effect, i.e., the decrease in $\rho_{zz}$ as a function of $B$.
Fig. \ref{fig_6}(b) shows the out-of-plane resistivity normalized by zero-field value ($\rho_{zz}/\rho_{zz}^0$) at 0.32 GPa.
The local maximum decreases as the temperature increases, whereas the slope of the negative magnetoresistance effect is
less sensitive to the temperature.
Compared to $\rho_{xx}$, very clear SdH oscillation superposed on the negative magnetoresistance was observed in high-field region.
With increments in pressure up to $P_{CDW2}$,
the quasilinear negative slope in high-field region was gradually suppressed, as shown in Fig. \ref{fig_6}(a), though local maximum still exists at $\sim 1$ T.
After $P$ passes through $P_{CDW2}$, the increase in $\rho_{zz}$ below 1 T was hardly observed at 1.6 K,
though it appeared as the temperature increases, as shown in Fig. \ref{fig_6}(c).
The origin of drastic change in the functional form of $\rho_{zz}/\rho_{zz}^0$ is unclear at the present stage.
In the normal metallic phase above $P_{CDW1}$, the increase in resistivity below 1 T and SdH oscillation again observed at 1.6 K.
As shown in Fig. \ref{fig_6}(d), the negative slope of $\rho_{zz}/\rho_{zz}^0$ at 3.7 GPa is almost identical to that at 0.32 GPa.

In conventional Drude model, $\rho_{zz}$ remains unchanged as a function of magnetic field since the current parallel to the field direction is free from Lorentz force.
An artificial negative longitudinal magnetoresistance effect is known to caused by an extrinsic mechanism called current jetting \cite{Arnold_2016, Reis_2016}, which becomes dominant in systems with large non-saturating $\rho_{xx}$
and high mobility.
In such a system with large anisotropy factor $\rho_{xx}/\rho_{zz}$,
the electric current in the sample becomes strongly inhomogeneous in high magnetic fields, and artificial negative longitudinal magnetoresistance can appear depending on
the geometry of the potential contact.
As for the present case, let we assume that $\rho_{zz}$ does not drastically change from the zero-field value, i.e., remains in order of 1 $\mu\Omega$ cm.
Then, $\rho_{xx}/\rho_{zz}$ is estimated to be in order of 1, since Fig. \ref{fig_4}(a) indicates that
$\rho_{xx}$ is also in order of 1 $\mu\Omega$ cm in the present magnetic field and pressure range.
Thus, this extrinsic mechanism is excluded from the possible origin.
We regard that the observed negative longitudinal magnetoresistance is arise from the intrinsic electronic structure.
The negative longitudinal magnetoresistance effect is often discussed in Dirac/Weyl systems,
which is called chiral anomaly \cite{Son_2013, Xiong_2015}.
In this mechanism, charge carriers are pumped from one Weyl point to another with opposite chiralities via the lowest Landau level in the presence of $B$ parallel to the electric field.
However, we observed negative contribution in all pressure from ambient pressure to 4 GPa,
and no apparent enhancement was observed above $P_{CDW1}$,
which is evident in comparison between Fig. \ref{fig_6}(b) and \ref{fig_6}(d).
This results suggest that the presence of the two-dimensional hollow Fermi surface,
in which possible Dirac dispersion has been proposed, cause little influence on the emergence of negative contribution.
On the other hand, recent theoretical study showed that negative magnetoresistance effect can arise from the warp of the Fermi surface from the perfect sphere,
regardless of the nontrivial band topology \cite{Awashima_2019}.
Thus, more details of the Fermi surfaces in both CDW phase and normal metallic phase under high pressure are indispensable to discuss the origin of negative longitudinal magnetoresistance.
Here, we also note that isostructural LaAu$_x$Sb$_2$, which shows CDW transitions at $\sim 97$ K at ambient pressure,
shows weak positive longitudinal magnetoresistance at ambient pressure up to 8 T (Fig. \ref{fig_s9} in Appendix \ref{app_Au}).
The sample size and configurations of electrical contact are almost identical with those in LaAgSb$_2$.
This supports that the remarkable negative longitudinal magnetoresistance is unique to the electronic structure of LaAgSb$_2$.

\begin{figure*}[]
\centering
\includegraphics[]{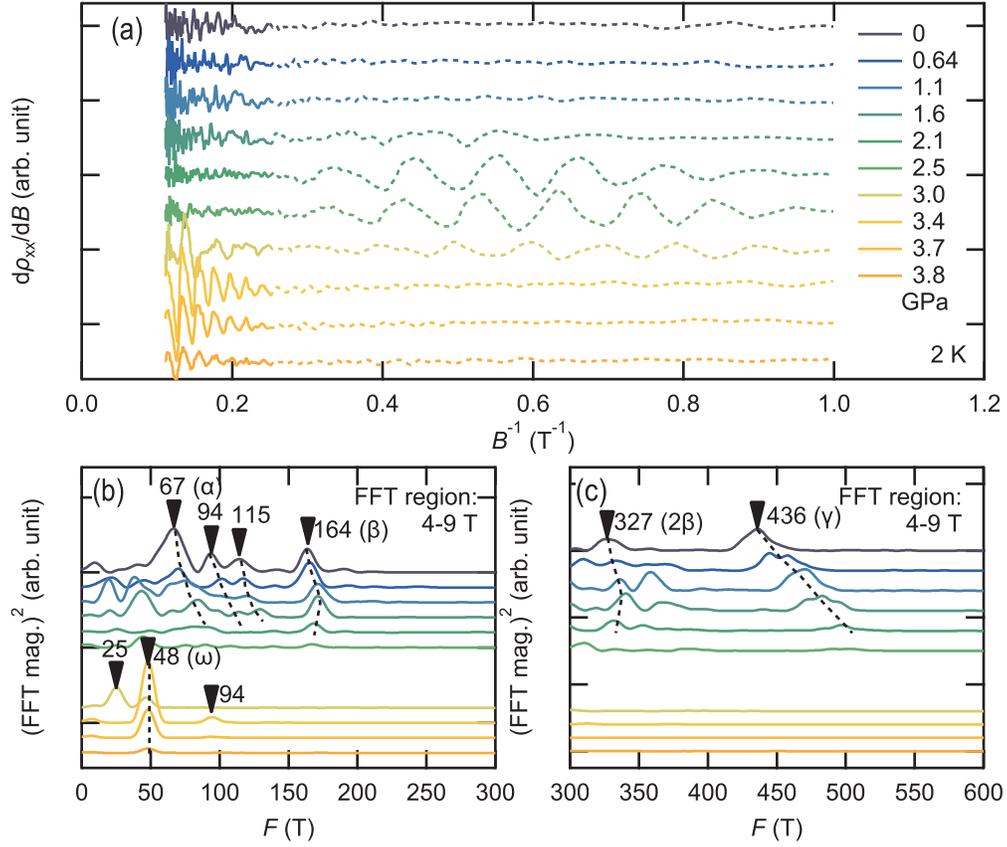}
\caption{
(a) Magnetic-field derivative of in-plane magnetoresistivity ($d\rho_{xx}/dB$) at 2 K as a function of $B^{-1}$.
The solid and broken region indicates the magnetic field of 4 T$<B<9$ T and 1 T$<B<4$ T, respectively.
FFT spectra of $d\rho_{xx}/dB$ (b) up to $F=300$ T and (c) above $F=300$ T.
In (b) and (c), the FFT was performed in 4 T$<B<9$ T.
Peaks are indicated by solid arrows with each frequency.
FFT spectra at 3.0, 3.4, 3.7, 3.8 GPa are multiplied by 0.1
to make uniform the height of peaks.
\label{fig_5}}
\end{figure*}
Then, we focus on the oscillatory structure on $\rho_{xx}$.
Figure \ref{fig_5}(a) shows the magnetic-field derivative of $\rho_{xx}$ at 2 K as a function of $B^{-1}$.
At ambient pressure, the oscillation is discernible above 4 T, and the oscillation pattern changes with application of pressure up to 3.8 GPa.
In particular, we can recognize a drastic change in oscillation pattern when $P$ passes through $P_{CDW1}$.
The oscillation above 4 T is periodic with respect to $B^{-1}$ and
damps as the temperature increases,
which is well described by conventional Lifshitz--Kosevich (LK) formula for quantum oscillation phenomena \cite{Shoenberg_MOM}.
Thus, this can be ascribed to Shubnikov--de Haas (SdH) oscillation by extremum cyclotron orbit of the Fermi surfaces.
Below 4 T (dotted region in Fig. \ref{fig_5}(a)), on the other hand,
we can also recognize another oscillatory structure with lower frequency between $P_{CDW2}$ and $P_{CDW1}$.
This oscillation is also assumed to be SdH oscillation,
though its amplitude shows anomalous temperature/magnetic-field dependence
unlike those above 4 T.
In the following, we firstly focus on the oscillation structure observed above 4 T, 
and the oscillation observed below 4 T will be discussed later.

\begin{table*}
\caption{\label{tab:SdH_summary}The oscillation frequency ($F$), cyclotron effective mass ($m_c$), Dingle temperature ($T_D$),
and pressure derivative normalized by ambient-pressure value ($1/F_0\cdot dF/dP$) obtained in the present study.
$F$ and $m_c$ are obtained from $\rho_{xx}$ data at ambient pressure unless otherwise specified by footnotes.
Data in the parentheses are taken from the previous studies.}
\begin{ruledtabular}
\begin{tabular}{lllll}
label& $F$ (T) & $m_c$ ($m_0$) & $T_D$ (K) & $1/F_0\cdot dF/dP$ (GPa$^{-1}$)\\
\hline
$\alpha$ & 67 (72 \cite{Myers_1999b})& $0.057\pm0.001$& & $0.16\pm0.04$\\
& 94& &&$0.135\pm0.001$\\
& 115& &&$0.07\pm0.02$\\
$\beta$ & 164 (164 \cite{Myers_1999b}, 160 \cite{Inada_2002})& $0.154\pm0.006$ (0.16 \cite{Myers_1999b}, 0.16 \cite{Inada_2002})&(0.67 \cite{Inada_2002})&$0.033\pm0.005$\\
$\gamma$ & 436 (432 \cite{Myers_1999b}, 440 \cite{Inada_2002})&$0.26\pm0.02$ \footnotemark[1] (0.28 \cite{Myers_1999b}, 0.34 \cite{Inada_2002})&(1.1 \cite{Inada_2002})&$0.081\pm0.004$ (0.16 \cite{Myers_1999b})\\
$\omega$ & 48 \footnotemark[2]&$0.066\pm0.001$ \footnotemark[2]&$34\pm2 \footnotemark[2]$&$0.04\pm0.01$\\
\end{tabular}
\end{ruledtabular}
\footnotetext[1]{Data obtained from $\rho_{zz}$ at 0.32 GPa.}
\footnotetext[2]{Data obtained from $\rho_{xx}$ at 3.4 GPa.}
\end{table*}

Figure \ref{fig_5}(b) and (c) shows the fast Fourier transform (FFT) spectra
of $d\rho_{xx}/dB$ at various pressure.
The FFT was performed in the magnetic field region between 4 and 9 T.
At ambient pressure, we identified six peaks, as indicated by arrows in Fig. \ref{fig_5}(b) and (c).
Following the previous quantum oscillation studies \cite{Myers_1999b, Inada_2002, Budko_2008},
the peaks with frequencies $F= 67, 164$, and 436 T are labeled as $\alpha$, $\beta$, and $\gamma$ peaks,
respectively.
These peaks shows quite well accordance with the previous studies as shown in Tab. \ref{tab:SdH_summary}.
The peak at 327 T is assigned to be second harmonic of $\beta$ since the pressure dependence of the frequency is identical with that of $\beta$.
The adopted temperature and magnetic-field range are different from those in previous studies,
and thus, oscillatory component higher than 600 T was not identified in the present study.
In the pressure region below $P_{CDW2}$, we recognized several FFT peaks below 60 T,
though it was difficult to trace the reliable pressure dependence.
As the pressure approaches to $P_{CDW2}$, all these peaks got gradually weaker and we could not trace the pressure dependence above 2.5 GPa.
Then, above $P_{CDW1}$, a single oscillation with frequency of 48 T becomes dominant.
At 3.0 and 3.4 GPa, just above $P_{CDW1}$, secondary oscillation components with frequencies of 25 and 94 were also identified.

\begin{figure}[]
\centering
\includegraphics[]{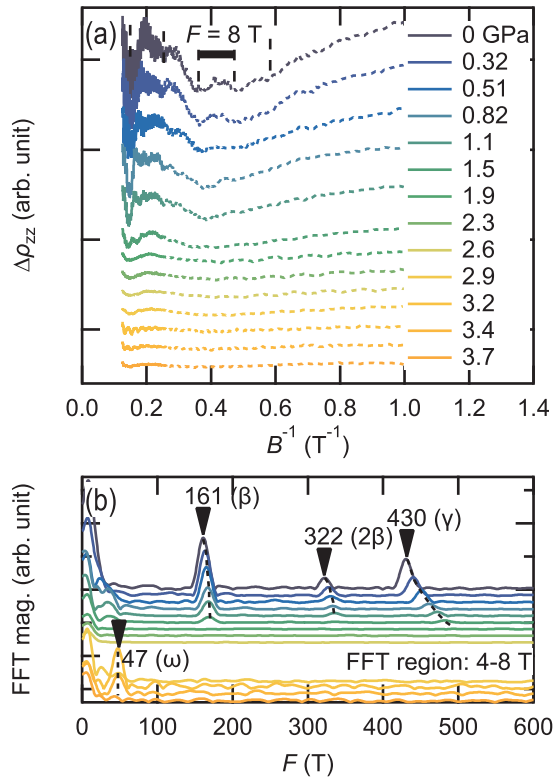}
\caption{
(a) Oscillatory component of the out-of-plane magnetoresistivity ($\Delta \rho_{zz}$) at 2 K as a function of $B^{-1}$.
$\Delta \rho_{zz}$ was obtained by subtraction of forth-polynomial background from
$\rho_{zz}$.
The solid and broken region indicates the magnetic field of 4 T$<B<8$ T and 1 T$<B<4$ T, respectively.
(b) FFT spectra of $\Delta \rho_{zz}$ obtained from magnetic-field region of 4 T$<B<8$ T.
Identified peaks are indicated by solid arrows with each frequency.
FFT spectra at 2.9, 3.2, 3.4, 3.7 GPa are multiplied by 5
to make uniform the height of peaks
\label{fig_7}}
\end{figure}
Then, we focus on the quantum oscillation in $\rho_{zz}$.
Figure \ref{fig_7}(a) shows the out-of-plane magnetoresistivity with fourth-polynomial background subtraction.
Compared with the in-plane magnetotransport, 
oscillation structure is quite sharp in the field region above 4 T,
and another oscillation component below 4 T also appeared in the pressure region $P_{CDW2}<P<P_{CDW1}$.
Figure \ref{fig_7}(b) shows the FFT spectra of SdH oscillation above 4 T.
We identified $\beta$, $2\beta$, and $\gamma$ peaks at ambient pressure, whose frequency agree quite well with those in
in-plane magnetotransport.
We also identified a oscillatory component with small frequency of $\sim 8$ T
[indicated by vertical broken lines and frequency scale in Fig. \ref{fig_7}(a)],
which has not been reported in previous studies.
As seen in Fig. \ref{fig_7}(b), the FFT peaks become gradually weak, and we could not see any discernible peak between $P_{CDW2}$ and $P_{CDW1}$.
Above $P_{CDW1}$, single peak with $F=47$ T appeared, which also consistent with Fig. \ref{fig_5}

\begin{figure}[]
\centering
\includegraphics[]{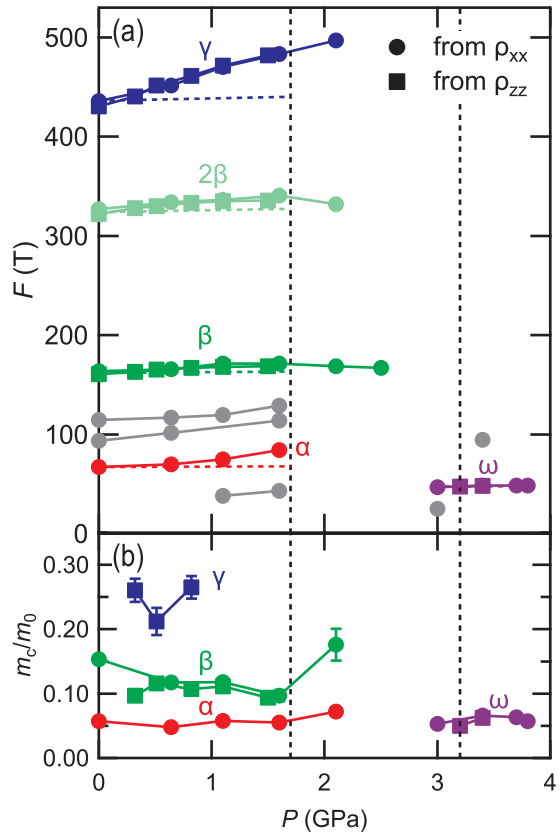}
\caption{
(a) Pressure dependence of the frequency of SdH oscillation ($F$) for $\alpha$, $\beta$, $\gamma$, and $\omega$ branches.
Approximately horizontal broken lines indicate the pressure dependence of $F$
expected from the compression of in-plane lattice constant.
(b) Pressure dependence of the cyclotron effective mass ($m_c$) for $\alpha$, $\beta$, $\gamma$, and $\omega$ branches.
The circular and rectangular markers indicate the data obtained from $\rho_{xx}$ and $\rho_{zz}$ measurements, respectively.
\label{fig_8}}
\end{figure}
The pressure dependence of the FFT peaks obtained from $\rho_{xx}$ and $\rho_{zz}$ measurements are summarized in Fig. \ref{fig_8}(a).
$\alpha$, $\beta$, $\gamma$, and $\omega$ shift almost linearly with the increment in pressure,
while only $\beta$ showed tendency of saturation as it approaches $P_{CDW2}$
and slight decrement in $P_{CDW2}<P<P_{CDW1}$.
In general, the cross section of a Fermi surface becomes larger as the first Brillouin zone is enlarged by shrinkage of lattice constant.
Using the linear compressibility of in-plane lattice parameter $1/a_0\cdot da/dP=-2.9\times 10^{-3}$ GPa$^{-1}$ \cite{Budko_2006},
the pressure derivative of the SdH frequency due to the shrinkage of lattice constant is estimated as $1/F_0\cdot dF/dP=5.8\times 10^{-3}$ GPa$^{-1}$.
Here, $a_0$ and $F_0$ represents the in-plane lattice constant and SdH frequency at ambient pressure.
Pressure dependence of lattice constant is approximated to be linear for simplicity.
We estimate $1/F_0\cdot dF/dP$ for each branch
by linear curve fitting in the pressure range of $0<P<P_{CDW2}$,
which is listed in Tab. \ref{tab:SdH_summary}.
For all branches, $1/F_0\cdot dF/dP$ is significantly larger than that expected from the lattice compression,
indicating that the change in the cross sections under pressure is dominated by modification of the band structure
that is not scaled by lattice deformation.
For comparison, the pressure dependence of $F$ for each branch expected from the lattice deformation is shown by approximately horizontal broken line in Fig. \ref{fig_8}(a).
This suggests that the deformation effect of the Fermi surface cannot be ignored to discuss the electronic structure under pressure.
For $\beta$ branch, 
a previous study has been estimated the pressure derivative 0.16 GPa$^{-1}$ \cite{Myers_1999b}.
The disagreement with the present study may come from limited data point and pressure region in the previous study.
We also estimated the cyclotron effective mass ($m_c$) from the temperature dependence of SdH oscillations for primary branches,
the result of which is summarized in  Fig. \ref{fig_8}(b).
$m_c$ of $\alpha$, $\beta$, and $\gamma$ branches at ambient pressure were confirmed to be consistent with previous quantum oscillation measurements,
as described in Tab. \ref{tab:SdH_summary}.
$m_c$ is not so sensitive to the pressure for all branches below $P_{CDW2}$,
and the $\omega$ branch in the normal metallic phase
above $P_{CDW1}$ holds small effective mass of $\sim 0.066$ $m_0$.
As we can see in Fig. \ref{fig_8}, the SdH oscillation observed in $\rho_{xx}$ and $\rho_{zz}$ shows quite good agreement with each other.
Although all Fermi surfaces should contribute to the transport properties above $P_{CDW1}$,
we found only $\omega$ peaks in the $B$ and $T$ range of the present study,
possibly due to larger external cross section and heavier cyclotron effective masses of the other orbits compared to $\omega$.

\begin{figure}[]
\centering
\includegraphics[]{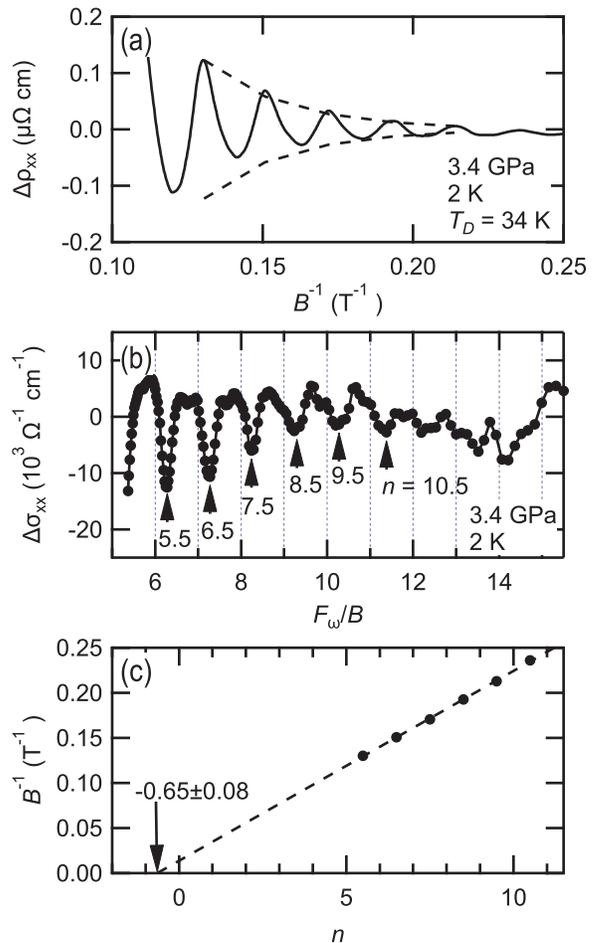}
\caption{
(a) Oscillatory component of in-plane magnetoresistivity ($\Delta \rho_{xx}$) at $P=3.4$ GPa and $T=2$ K as a function of $B^{-1}$.
The dotted curve represents the envelope function reproduced by Dingle temperature of $T_D=34$ K (see main text).
(b) Oscillatory component of in-plane conductivity ($\Delta \sigma_{xx}$) at $P=3.4$ GPa and $T=2$ K as a function of $F_\omega/B$.
$F_\omega=48$ T represents the FFT frequency obtained at 3.4 GPa.
Fourth-polynomial background was subtracted from the original $\Delta \sigma_{xx}$.
Dip positions used to construct (c) are indicated by solid arrows.
(c) Inverse magnetic field of the dip positions as a function of Landau index $n$.
The dip positions were assigned to the half-integer $n$.
\label{fan}}
\end{figure}
We further focus on the $\omega$ branch in the normal metallic phase under pressure.
The corresponding cross section in the $\bm{k}$ space ($S_\omega$) is estimated to be $S_\omega=2\pi e F_\omega /\hbar = 4.58\times 10^{17}$
m$^{-2}$ with using $F_\omega=48$ T at 3.4 GPa.
Comparing with the cross section of the first Brillouin zone at 3.4 GPa, $S_{BZ}=2.09\times 10^{20}$ m$^{-2}$,
$S_\omega$ is only 0.22\% of $S_{BZ}$, indicating a quite small cyclotron orbit.
Based on the Fermi surface at ambient pressure \cite{Myers_1999b, Inada_2002},
the small ellipsoidal electron pocket located at the $X$ point might be the origin of this frequency.
However, whether the Fermi surface at ambient pressure is directly applicable to the case under pressure is unclear at present,
considering the unignorable band deformation suggested in the pressure dependence of the SdH frequency.
Further investigation of field-angular dependence of this frequency
and careful comparison with band calculation under pressure are needed to clarify the geometry of the Fermi surface in the normal metallic phase.

Figure \ref{fan}(a) shows $\Delta \rho_{xx}$ at $P=3.4$ GPa and $T=2$ K as a function of $B^{-1}$.
The SdH oscillation originating from $F_\omega$ shows a clear bell-bottom-shaped envelope,
which enables the Dingle temperature $T_D$ to be deduced.
According to the LK formula, the field dependence of the amplitude $A(B)$ is expected as follows, with cyclotron frequency $\omega_c=eB/m_c$: \cite{Shoenberg_MOM}
\begin{equation}
A(B)\propto B^{-1/2} \dfrac{\exp[-2\pi^2k_B T_D / (\hbar \omega_c)]}{\sinh[2\pi^2k_B T / (\hbar \omega_c)]}.
\label{eq_env}
\end{equation}
From the plot of $\ln \{A(B)B^{1/2}\sinh[2\pi^2k_B T / (\hbar \omega_c)] \}$ as a function of $B^{-1}$,
$T_D=34\pm2$ K was extracted.
Here, we adopted $m_c=0.066$ $m_0$, estimated from the temperature dependence of the amplitude at 3.4 GPa.
As shown by the dotted curves in Fig. \ref{fan}(a), the envelope function was reproduced well by Eq. (\ref{eq_env}), assuming the obtained $T_D$.
Correspondingly, the relaxation time $\tau_D=\hbar/(2\pi k_B T_D)$ and mobility $\mu_D=e\tau_D/m_c$
were estimated as $\tau_D=(3.6\pm0.2)\times 10^{-14}$ s and $\mu=950\pm 60$ cm$^2$/(Vs), respectively.

To obtain the insight on possible existence of nontrivial band topology in the normal metallic phase above $P_{CDW1}$,
we analyzed the phase factor of the SdH oscillation in the in-plane conductivity $\sigma_{xx}$.
Figure \ref{fan}(b) shows the oscillatory component of the in-plane conductivity ($\Delta \sigma_{xx}$)
at $P=3.4$ GPa and $T=2$ K as a function of $F_\omega/B$.
Here, $\Delta \sigma_{xx}$ was obtained by subtraction of the fourth-polynomial background from $\sigma_{xx}$.
The dips in $\Delta \sigma_{xx}$ were quite sharp, as indicated by the solid arrows in Fig. \ref{fan}(b),
whereas the peaks of $\Delta \sigma_{xx}$, which are ascribed to level-crossing points, exhibited relatively flat figures.
We associated the inverse magnetic field of the dip position to the half-integer Landau index,
whose relationship is shown in Fig. \ref{fan}(c).
The horizontal intercept of this Landau-level fan diagram is discussed to identify the anomalous phase shift originating
from the presence of the Dirac fermion \cite{Ando_2013, Murakawa_2013}.
The functional form of $\Delta \sigma_{xx}$ is simplified as 
\begin{equation}
\Delta \sigma_{xx}\propto \cos \left[ 2\pi \left(\dfrac{F_\omega}{B}+\gamma \pm \delta \right)\right].
\end{equation}
$\gamma=1/2-\Phi_B/(2\pi)$ is believed to reflect the Berry phase $\Phi_B$, which is $\pi$ for a Dirac system and zero for a trivial system.
$\delta$ depends on the dimensionality of the system: $\delta=\pm1/8$ for the three-dimensional case and $\delta=0$ for the two-dimensional case.
The above formulation indicates that in the three-dimensional system,
the horizontal intercept of the Landau-level fan diagram takes $-1/2\pm1/8$ and $\pm1/8$
for the trivial and Dirac case, respectively.
In the case of the $\omega$ branch, the horizontal intercept is determined to be $-0.65\pm0.08$, which seems to be close to the trivial case ($-1/2-1/8\sim-0.625$).
However, we note that the phase of quantum oscillation is considerably affected by whether Zeeman splitting exists or not \cite{Akiba_2018}.
The effect of Zeeman splitting yields an amplitude factor
$R_s=\cos[\pi g^*\mu_B B/(\hbar \omega_c)]=\cos(\pi M_{ZC})$,
whose sign changes depending on the effective $g$ factor ($g^*$)
and cyclotron effective mass.
The flat peaks in Fig. \ref{fan}(b) may indicate that this branch possesses finite Zeeman splitting, which is inconspicuous, owing to the limited magnetic field range
in the present study.
Thus, to properly evaluate the topological aspect from SdH oscillation, 
the value of the Zeeman--cyclotron ratio $M_{ZC}$,
which is an alternative index to discuss the nontrivial band topology \cite{Fuseya_2015, Hayasaka_2016},
should be clarified by further
high-field or field-angular dependence measurements.
The horizontal intercept of $\omega$ branch determined in the present study suggests $2N-0.5<M_{ZC}<2N+0.5$, where $N=0, 1, 2, \dots$.
The specification of $M_{ZC}$ remains for future study.
\begin{figure*}[]
\centering
\includegraphics[]{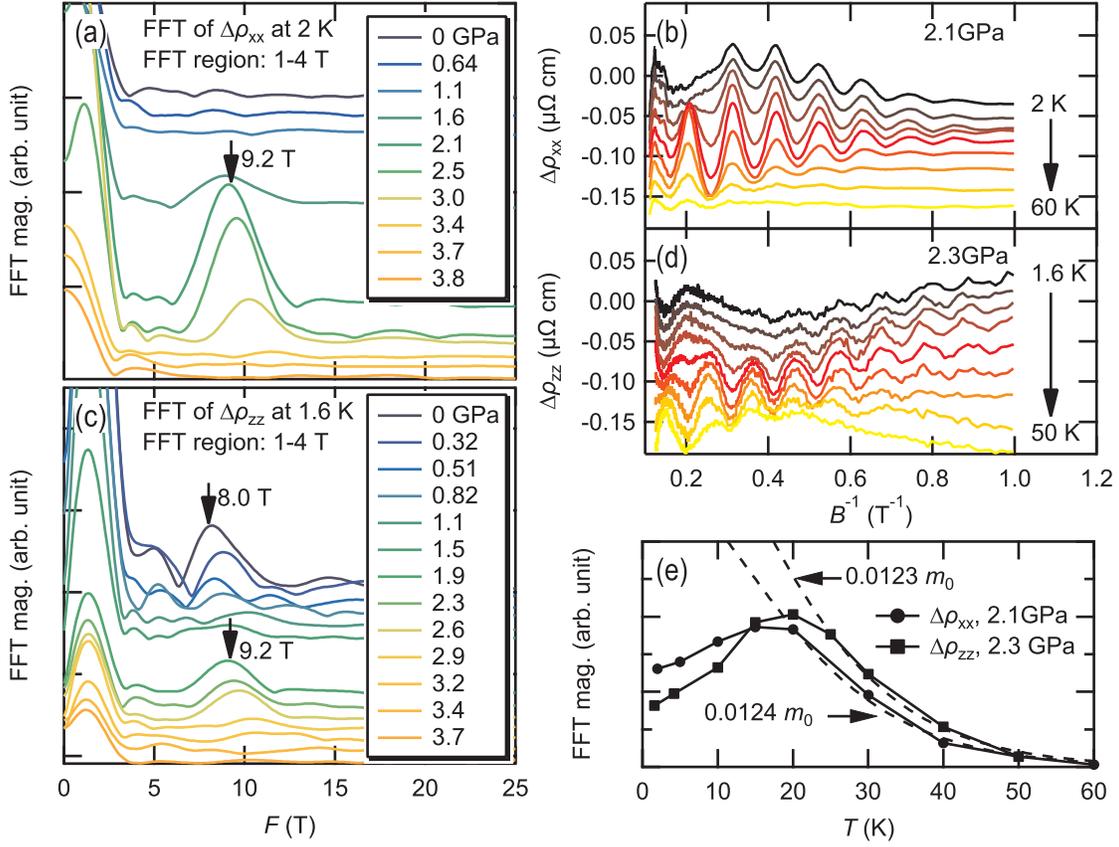}
\caption{
FFT spectra of oscillatory component of (a) the in-plane magnetoresistivity ($\Delta \rho _{xx}$) at 2 K and (c) out-of-plane magnetoresistivity ($\Delta \rho _{zz}$) at 1.6 K.
In (a) and (c), FFT analysis were performed in the magnetic-field region of 1 T$<B<4$ T.
(b) $\Delta \rho _{xx}$ at 2.1 GPa and (d) $\Delta \rho _{zz}$ at 2.3 GPa as a function of $B^{-1}$.
Each curve is vertically shifted for clarity.
(e) Temperature dependence of the FFT magnitude for $\Delta \rho _{xx}$ at 2.1 GPa and $\Delta \rho _{zz}$ at 2.3 GPa.
FFT analysis were performed in the magnetic-field region of 1 T$<B<4$ T.
\label{fig_9}}
\end{figure*}

Finally, we focus on the oscillatory component observed below 4 T [dotted area in Fig. \ref{fig_5}(a) and Fig. \ref{fig_7}(a)]
in the pressure range of $P_{CDW2}<P<P_{CDW1}$.
Figure \ref{fig_9}(a) shows the FFT spectra of $\Delta \rho_{xx}$ between 1 and 4 T at 2 K.
A FFT peak of $\sim 9.2$ T is discernible at 1.6 GPa where CDW2 still exists, and it was suddenly enhanced in $P_{CDW2}<P<P_{CDW1}$.
As $P$ increases toward $P_{CDW1}$, the oscillation becomes considerably damped and is not observed above $P_{CDW1}$.
Figure \ref{fig_9}(b) shows the temperature variation of $\Delta \rho_{xx}$ at 2.1 GPa, where the FFT magnitude at 2 K reaches its maximum.
We can see that the peak at $B^{-1}=0.2$ T$^{-1}$ shows anomalous behavior:
although it is absent below 10 K, it becomes large as the temperature increases, and then decreases.
It seems that the oscillation amplitude at high-$B$ and low-$T$ region is strongly suppressed for some reason.
The oscillation with frequency of $\sim 9.2$ T was also reproduced in the out-of-plane resistivity measurement.
Figure \ref{fig_9}(c) shows the FFT spectra of $\Delta \rho_{zz}$ between 1 to 4 T at 1.6 K.
We can recognize the enhancement of the FFT peak with $F\sim 9.2$ T in $P_{CDW2}<P<P_{CDW1}$,
which is consistent with the result shown in Fig. \ref{fig_9}(a).
We can also identify a peak with $F\sim 8$ T at ambient pressure, which is considered to be a long-period oscillation, as described in Fig. \ref{fig_7}(a).
Whether the oscillation with $F\sim 8$ T at ambient pressure has an identical origin with that observed in $P_{CDW2}<P<P_{CDW1}$ remains unclear at present,
although the frequencies are close with each other.
Figure \ref{fig_9}(d) shows the temperature variation of $\Delta \rho_{zz}$ at 2.3 GPa.
The suppression of oscillation amplitude at the high-$B$ and low-$T$ region is seen more clearly.

The distinct peak in the FFT spectrum indicates that this oscillation is well periodic as a function of $B^{-1}$, suggesting a sort of magnetotransport effect.
Besides, the frequency is independent of current direction as long as the magnetic-field direction is identical, which agree with the general feature of the SdH oscillation.
However, the significant reduction of the amplitude observed at the high-$B$ and low-$T$ is unusual compared with conventional SdH oscillation.
Several mechanisms have been known to cause the resistivity oscillation,
whose amplitude shows anomalous temperature dependence
although they are quite similar with the SdH oscillation.
One is called magneto-phonon resonance \cite{Nicholas_1985},
whose origin is resonant carrier scattering across the Landau levels caused by optical phonon.
However, this phenomenon is mostly observed in low-carrier systems such as semiconductors and usually absent below 50 K, where the population of the optical phonon is quite small.
These features do not agree with our results.
Another example is called Stark quantum interference \cite{Shoenberg_MOM, Stark_1971}.
The origin of this resistivity oscillation is a quantum interference of the wavefunction between two open orbits running parallel in the momentum space.
However, the oscillation amplitude is expected to be almost independent of temperature unless the orbits are considerably smeared out.
Thus, the strong temperature dependence of the resistivity peak at high-$B$ region cannot be explained by this mechanism.
Thus, we ascribe the SdH oscillation being most probable origin of this oscillation phenomenon. 
The deviation from the conventional LK formula may caused by a change of $m_c$ and/or $g^*$, though the precise origin is unclear within the scope of the present study.
The small frequency of 9.2 T corresponds to the 
cross section of $8.78\times 10^{16}$ m$^{-2}$ in the $\bm{k}$ space,
which is only 0.042\% of the cross section of the first Brillouin zone at 2.1 GPa.
Because such a small cross section is not expected in the band calculation at ambient pressure,
the emergence of small Fermi surface under pressure (Lifshitz transition) or reconstruction of Fermi surface by nesting of CDW1 is considered to be possible origin.

Figure \ref{fig_9}(e) summarizes the temperature dependence of the FFT magnitude calculated from the data shown in Fig. \ref{fig_9}(b) and \ref{fig_9}(d).
The temperature dependence of the oscillation showed almost identical property between $\Delta \rho_{xx}$ and $\Delta \rho_{zz}$.
As shown in the broken lines in Fig. \ref{fig_9}(e), the
temperature dependence of the FFT magnitude above 20 K is reproduced well by conventional the LK formula with fixed $m_c\sim 0.012$ $m_0$.
The reduction of the FFT magnitude below 20 K reflects the damping behavior described in Figs. \ref{fig_9}(b) and \ref{fig_9}(b).

\begin{figure}[]
\centering
\includegraphics[]{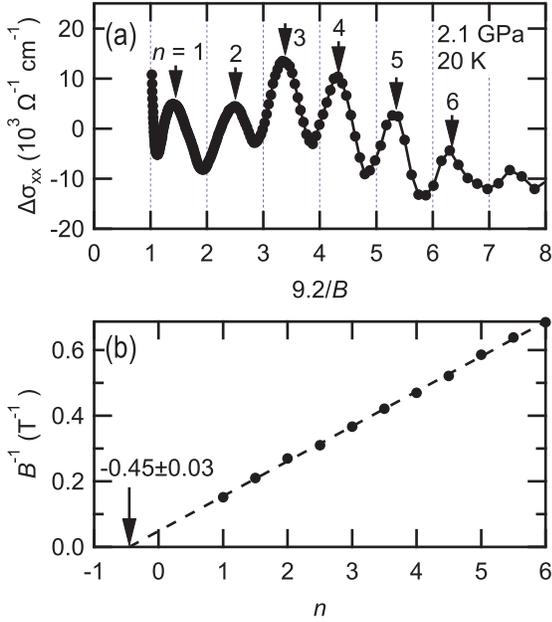}
\caption{
(a) Oscillatory component of in-plane conductivity $\Delta \sigma_{xx}$ at 2.1 GPa and 20 K.
Fifth-polynomial background was subtracted from $\sigma_{xx}$.
Horizontal axis represents the inverse magnetic field normalized by the frequency of 9.2 T.
(b) Landau-level fan diagram constructed from the data in (a).
The integer $n$ corresponds to the peak in $\Delta \sigma_{xx}$.
\label{fig_10}}
\end{figure}
At last, we focus on the phase factor of this small cyclotron orbit.
Figure \ref{fig_10}(a) shows the oscillatory component of conductivity ($\Delta \sigma_{xx}$) at 2.1 GPa and 20 K.
As well as the case of $\omega$ branch, we assigned the peak position to the
integer Landau index, and constructed Landau-level fan diagram is shown in
Fig. \ref{fig_10}(b).
The horizontal intercept was estimated to be $-0.45\pm0.03$, which seems to be close to
the trivial case ($-1/2+1/8\sim-0.375$),
or $2N-0.5<M_{ZC}<2N+0.5$ where $N=0, 1, 2, \dots$ using the Zeeman--cyclotron ratio.

\section{Conclusion}
In conclusion, we comprehensively investigated the magnetotransport properties of 
LaAgSb$_2$ under high pressure to clarify the phase diagram of the two charge-density-wave (CDW) states
and the change in the electronic structure across the CDW critical pressures.
The temperature dependence of in-plane, out-of-plane, and Hall resistivity showed
anomalies at $T_{CDW1}\sim 210$ K and $T_{CDW2}\sim 190$ K at ambient pressure; this result agrees with the previous studies.
Both $T_{CDW1}$ and $T_{CDW2}$ were suppressed under pressure with rates of
$-51$ and $-80$ K/GPa, respectively.
Compared with LaAuSb$_2$, which shows similar successive CDW transitions,
the CDW states in LaAgSb$_2$ seems robust against pressure.
The critical pressure of CDW1 and CDW2 were determined to be $P_{CDW1}\sim 3.0$--3.4 GPa and $P_{CDW2}\sim 1.5$--1.9 GPa, respectively.

The $\rho_{xx}$ measurements ($B\parallel c$, $I\perp B$)
under pressure showed that
the quasilinear magnetoresistance effect (1000\% in $P<P_{CDW2}$ and $P>P_{CDW1}$, and 4000\% in $P_{CDW2}<P<P_{CDW1}$ at 9 T).
In-plane Hall conductivity showed step-like increase at $P_{CDW1}$.
This is qualitatively consistent with the emergence of two-dimensional hollow Fermi surface,
which is regarded to responsible for the nesting of CDW1.
$\rho_{zz}$ ($B\parallel I \parallel c$) showed prominent negative magnetoresistance effect, which was observed to some extent in all measured pressures from ambient to 4 GPa.
The presence of hollow Fermi surface seems to be insensitive to the emergence of
this negative magnetoresistance effect.
Additional information about the Fermi surface under pressure
is necessary to understand the field dependence of $\rho_{zz}$. 

We observed clear Shubnikov--de Haas (SdH) oscillation
in both $\rho_{xx}$ and $\rho_{zz}$, whose oscillation frequencies were consistent with those found in the previous quantum oscillation studies at ambient pressure.
The pressure dependence of the oscillation frequencies were significantly larger than those expected
from the decrease in the in-plane lattice constant,
which indicates unignorable deformation of the Fermi surface
under pressure.
In the normal metallic state above $P>P_{CDW1}$,
we observed a single SdH oscillation with a frequency and cyclotron mass of
48 T and 0.066 $m_0$, respectively.
This cyclotron orbit corresponds to only 0.22\% of the first-Brillouin-zone area projected on $k_z=0$.
The Landau-level fan diagram analysis suggested the trivial phase factor for this oscillation.
In the pressure region in the present study ($P<4$ GPa),
the cyclotron effective mass for each oscillation component showed
little pressure dependence.
We also identified another oscillatory component below 4 T,
which is significantly enhanced for $P_{CDW2}<P<P_{CDW1}$.
The oscillation was periodic in terms of the inverse magnetic field
and showed an identical frequency of 9.2 T in $\rho_{xx}$ and $\rho_{zz}$.
We ascribe that this is an SdH oscillation from its small cyclotron orbit
(0.042\% of the first Brillouin zone).
The amplitude of this SdH oscillation was anomalously suppressed in the low-temperature and high-field region, and its origin has been left for future study.

\begin{acknowledgments}
We thank S. Araki and H. Harima for their many helpful comments and discussions, H. Ota for the X-ray single-crystal structural analyses, and M. Yokoyama for support with Laue diffraction measurements. This research was supported by JSPS KAKENHI Grant Number 19K14660. X-ray single-crystal structural analyses were performed at the Division of Instrumental Analysis, Okayama University. 
\end{acknowledgments}

\bibliography{reference}

\clearpage

\appendix

\section{Electrical transport properties of LaAu$_x$Sb$_2$}
\label{app_Au}
\begin{figure}[]
\centering
\includegraphics[]{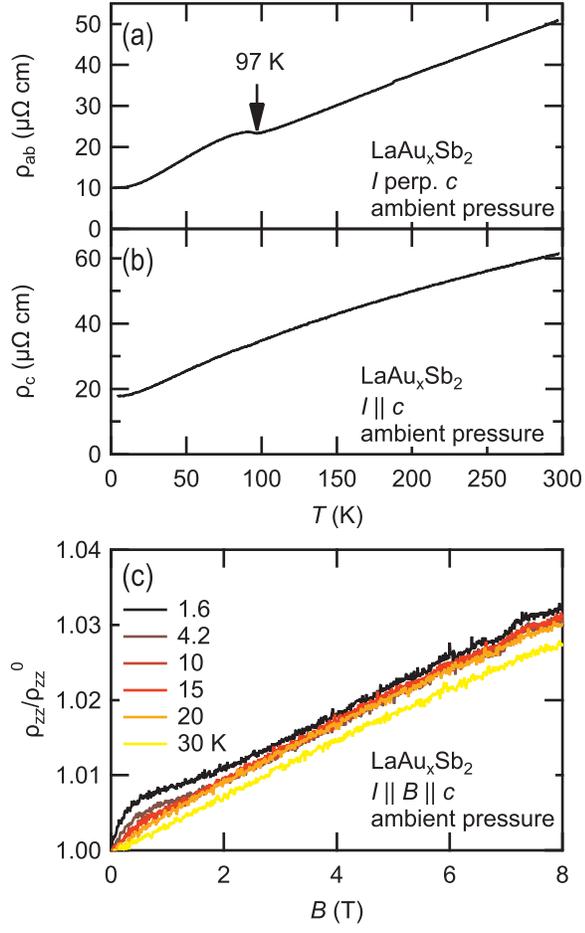}
\caption{
Temperature dependence of the (a) in-plane resistivity and (b) out-of-plane resistivity in LaAu$_x$Sb$_2$.
(c) Out-of-plane magnetoresistivity ($I\parallel B \parallel c$) normalized by zero-field value at ambient pressure.
\label{fig_s9}}
\end{figure}

Single crystals of LaAu$_x$Sb$_2$ were obtained by Sb self-flux method \cite{Kuo_2019}.
La (99.9\%), Au (99.9\%), and Sb (99.9999\%) with a molar ratio of 1:2:20 were placed in an alumina crucible,
and synthesized by the same process described in LaAgSb$_2$.
The temperature dependence of in-plane resistivity ($\rho_{ab}$) at ambient pressure is shown in Fig. \ref{fig_s9}(a).
We observed a hump-like anomaly at $T_{CDW1}=97$ K, which is ascribed to the CDW transition with higher transition temperature.
Based on Ref. \cite{Xiang_2020}, the amount $x$ is affected by
initial molar ratio of the elements.
Although we did not evaluate the precise amount of Au,
$T_{CDW1}$ of our sample seems to be close to the case of LaAu$_{0.947}$Sb$_2$
in Ref. \cite{Xiang_2020}.
We could not identify the secondary CDW transition with lower transition temperature, possibly owing to the off-stoichiometric nature.
Figure \ref{fig_s9}(b) shows the out-of-plane resistivity ($\rho_{c}$) at ambient pressure.
The anomaly at $T_{CDW1}$ is quite weak compared to $\rho_{ab}$,
which is consistent with the previous report \cite{Xiang_2020}.
Figure \ref{fig_s9}(c) shows the longitudinal magnetoresistivity
normalized by zero-field value ($\rho_{zz}/\rho_{zz}^0$) at ambient pressure.
The magnetoresistance effect is positive and quite small, which contrasts with the
prominent negative magnetoresistance effect observed in LaAgSb$_2$.
We note that the sample size and geometry of electrical contact is almost identical to
those in LaAgSb$_2$.

\end{document}